\documentclass[oneside,onecolumn]{article}

\usepackage{blindtext} 
\usepackage{graphicx}
\usepackage[sc]{mathpazo} 
\usepackage[T1]{fontenc} 
\usepackage{microtype} 

\usepackage[english]{babel} 

\usepackage[hmarginratio=1:1,top=32mm,columnsep=20pt]{geometry} 
\usepackage[hang, small,labelfont=bf,up,textfont=it,up]{caption} 
\usepackage{booktabs} 

\usepackage{lettrine} 

\usepackage{enumitem} 
\setlist[itemize]{noitemsep} 

\usepackage{abstract} 

\usepackage{titlesec} 
\renewcommand\thesection{\Roman{section}} 
\renewcommand\thesubsection{\roman{subsection}} 
\titleformat{\section}[block]{\large\scshape\centering}{\thesection.}{1em}{} 
\titleformat{\subsection}[block]{\large}{\thesubsection.}{1em}{} 

\usepackage{fancyhdr} 
\usepackage{titling} 

\usepackage{hyperref} 
\usepackage{breakurl}

\def\vec#1{\mathchoice{\mbox{\boldmath$\displaystyle#1$}}
{\mbox{\boldmath$\textstyle#1$}}
{\mbox{\boldmath$\scriptstyle#1$}}
{\mbox{\boldmath$\scriptscriptstyle#1$}}}

\pretitle{\begin{center}\Huge\bfseries} 
\posttitle{\end{center}} 
\title{Home advantage of European major football leagues under COVID-19 pandemic} 
\author{%
\textsc{Eiji Konaka}
 \\[1ex] 
\normalsize Meijo University \\ 
\normalsize \href{mailto:konaka@meijo-u.ac.jp}{konaka@meijo-u.ac.jp} 
}
\date{\today} 
\renewcommand{\maketitlehookd}{
\begin{abstract}
Since March 2020, the environment surrounding football has changed dramatically because of the COVID-19 pandemic.
After a few months' break, re-scheduled matches were held behind closed doors without spectators.
The main objective of this study is a quantitative evaluation of ``crowd effects'' on home advantage, using the results of these closed matches.
The proposed analysis uses pairwise comparison method to reduce the effects caused by the unbalanced schedule.
The following conclusions were drawn from the statistical hypothesis tests conducted in this study: In four major European leagues, the home advantage is reduced in closed matches compared to than in the normal situation, i.e.,  with spectators.
The reduction amounts among leagues were different.
For example, in Germany, the home advantage was negative during the closed-match period.
On the
other hand, in England, statistically significant differences in home advantage were not observed between closed matches and normal situation.

\end{abstract}
}

\begin{document}

\maketitle


\section{Introduction}

Sice March 2020, the environment surrounding football has changed dramatically --- because of the COVID-19 pandemic.
Similar to many other crowd-pleasing events, most football leagues were suspended.
Some of the leagues had even been calcelled, whereas others resumed after a few months' break. 
With regard to the resumed leagues, however, no decision as before the suspension, has been announced. Re-scheduled matches have therefore been held behind closed doors without spectators.

The objective of this study was to analyze how this unfortunate closed-match situation affected the match outcomes of football.
In particular, our main objective was a quantitative evaluation of ``crowd effects'' on home advantage.

Today, the existence of ``home advantage'' in sports, especially football, seems unquestionable \cite{nevill1999home}\cite{Pollard1986}\cite{Nevill1996}. 
However, definitive evidence on the factors that produce home advantage remains elusive. 
Of course, the quantitative impact of each of these factors on home advantage has also been discussed.

``Home advantage'' can be defined as the benefit that induces home teams to consistently win more than 50 percent of the games under a balanced home-and-away schedule \cite{Courneya1992}.
Quantitative research on home advantage in football dates back nearly 40 years to Morris \cite{Morris1981soccer}, followed by Dowie \cite{Dowie1982} and Pollard \cite{Pollard1986}.
In a very brief and thorough review paper on home advantage \cite{Pollard2008}, consisting of only three pages, Pollard presented eight main factors that could generate home advantage and explained conventional studies for each (of course, he did not forget to mention  interactions and other factors).
The first one is ``'crowd effect,'' i.e, effect caused by spectators.
In \cite{Pollard2008}, he wrote, ``This is the most obvious factor involved with home advantage and one that fans certainly believe to be dominant \cite{Wolfson2005}\cite{Lewis2007PerceptionsOC}.''
Studies by Dowie \cite{Dowie1982} and Pollard \cite{Pollard1986} were also able to find very little evidence that home advantage depends on crowd density (spectators per stadium capacity).

This study utilizes the results of matches conducted behind closed doors during the COVID-19 pandemic to determine the relationship between the presence of spectators and home advantage.
A similar study had already been performed by Reade et al., who used the results of matches conducted behind closed doors since 2003 and reported the results of their investigation on crowd effect \cite{https://www.carlsingletoneconomics.com/uploads/4/2/3/0/42306545/closeddoors_reade_singleton.pdf}.
The number of matches in empty studiums, however, were small (160, out of approximately 34 thousand matches). 
In addition, in these closed matches, teams were banned from admitting
supporters into their stadiums typically as punishments for bad behavior off the football pitch (e.g., bacause of corruption, racist abuse, or violence).
Therefore, in this analysis, the characteristics of the spectators could biased, e.g., they were excessively violent or exhibiting overly aggressive supporting behavior.

When the results of closed matches are used in such a study, it should also be noted that the schedule is unbalanced.
For example, many European league matches were already approximately two-thirds completeed by the mid-March suspension.
Therefore, a possible bias for the strength of home teams in the resumed matches after the break should be considered.
By contrast, most famous and extensive previous studies on home advantage \cite{Pollard2005} and the most recent report
\cite{https://www.economist.com/graphic-detail/2020/07/25/empty-stadiums-have-shrunk-football-teams-home-advantage} used only a number of basic statistics, e.g., numbe of goals, fouls, and wins.
These data could be biased if obtained under an unbalanced schedule.

For the problem on biased schedule, in this study, a statistical model that deremines match results based on the team strength parameters for each team and a home advantage parameter that is common for every team in the league was assumed.
Through the fitting the parameters in this model to minimize explanation error, the home advantage was separated even when the schedule was unbalanced.

The specific techniques used in this study were as follows: Each team $i$ had one strength evaluation value $r_i$, referred to as the``rating.'' 
Furthurmore, each league had one parameter, $r_{homeAdv}$, that expressed the home advantage.
The strength difference $\Delta r$ was defined as the difference in rating values between the teams, indexed by $i$ and $j$, added to the home advantage, i.e., $\Delta r=r_i+r_{homeAdv}-r_j$. The strength difference then explained the score ratio in a match via logistic regression model, i.e., $1/(1+\mathrm{exp}(-\Delta r))$.

The rating values $r_i$ and the home advantage value $r_{homeAdv}$ at a particular date were estimated using the most recent match results, e.g., for five matchweeks.
This calculation process was repeated for every matchweek. 
The home advantage values estimated only from the closed matches were then compared to those from past ``normal'' match results.

This paper is organized as follows:
Section \ref{sec:method} describes the data and the detailed algorithm used in this study.
An analysis on the five major top divisions of European football leagues, i.e., England, France, Germany, Italy, and Spain (in France, the top division has not resumed after suspension), is then presented. 
The match results were collected from 2010--2011 season.
Section \ref{sec:result} then discusses statistical analysis.
The following conclusions were able to be drawn from the statistical hypothesis tests  that were performed in this study.
\begin{itemize}
\item In the four major European leagues that were examined, the home advantage was reduced when there were no spectators compared to that for a normal situation,i.e., with spectators. 
\item The reduction amounts among the leagues were different. 
\item For all four leagues, the home advantage remained even in closed matches.
\end{itemize}
Lastly, Section \ref{sec:conclusion} summarizes and concludes this paper.

\section{Methods}
\label{sec:method}
In this section, the leagues that were investigated and the content of the used data  are described.
A mathematical method for estimating home advantage is then explained.

\subsection{Data set}

Table \ref{tab:data} outlines the leagues and the numbers of matches examined in this study. 
\begin{table}[htbp]
\begin{center}
\caption{Numbers of matches examined in this study}
\label{tab:data}
\begin{tabular}{llcccc}\hline
Country & League & Teams & Matches & 
	\multicolumn{2}{c}{Matches (2019/20)}   \\
 &  &  & (2010/11 --- 2018/19) & Normal & Closed \\ \hline 
England & Premier League & 20 & 3420 & 290 & 90 \\
France & Ligue 1 & 20 & 3420 & 279 & 0 \\
Germany & Bundesliga & 18 & 2754 & 216 & 90 \\
Italy & Serie A & 20 & 3420 & 240 & 140 \\
Spain & LaLiga & 20 & 3420 & 270 & 110 \\ \hline
Total &  &  & 16434 & 1295 & 430 \\ \hline
\end{tabular}
\end{center}
\end{table}

This study analyzed the home advantage among the top divisions in five European countries, i.e.,   England, France, Germany, Italy, and Spain, which are considered as the most major and highest-quality football leagues around the world.
The match results from the 2010/11 season were collected from worldfootball.net (\url{https://www.worldfootball.net/}). The number of matches analyzed was 17729, including 430 closed matches.

All five leagues were suspended from mid-March because of the COVID-19 pandemic. 
Four of the leagues, i.e., excluding France, resumed by late June, and finished by early August.
Ligue 1 in France, on the other hand, quickly decided and announced its cancellation at the end of April \cite{Ligue}.

Table \ref{tab:matchweek} summarizes their closed-match periods.
\begin{table}[htbp]
\begin{center}
\caption{Closed period}
\label{tab:matchweek}
\begin{tabular}{llccc} \hline
Country & League & Matchweeks & Closed from & Closed matchweeks \\ \hline
England & Premier League & 38 & 30 & 9 \\
Germany & Bundesliga & 34 & 25 & 10 \\
Italy & Serie A & 38 & 25 & 14 \\
Spain & LaLiga & 38 & 28 & 11 \\ \hline
\end{tabular}
\end{center}
\end{table}

\subsection{Mathematical model}
We propose a unified and simple statistical estimation method for scoring ratios based on the scores in each match, which are always officially recorded and are subject to a scoring system common to all the games.
This method extends \cite{Konaka2019IEICE-D} by incorporating home advantage. The study \cite{Konaka2019IEICE-D} reported that this proposed method achieved higher prediction accuracies for ten events of five sports, i.e., basketball, handball, hockey, volleyball, and water polo, in the Rio Olympic Games compared to those of official world rankings.

{
The scoring ratio of a home team $i$ in a match against an away team $j$ ($i$ and $j$ are team indices),  denoted as $p_{i,j}$, is estimated as follows:
\begin{equation}
\label{eqn:rateDiffToProb}
p_{i,j}=\frac{1}{1+e^{-(r_i+r_{homeAdv}-r_j)}},
\end{equation}
where  $r_i$ is defined as the {\it rating} of team $i$, and $r_{homeAdv}$ is the quantitative value of home advantage.

Given $(s_i, s_j)$, the actual scores in a match between $i$ and $j$,
\begin{equation}
s_{i,j}=\frac{s_i+1}{s_i+1+s_j+1}=p_{i,j}+\epsilon_{i,j},
\end{equation}
where $s_{i,j}$ and $\epsilon_{i,j}$ are the modified actual scoring ratio and the estimation error, respectively.
}
In football, shut-out results such as $1-0$ or $3-0$ occur frequently.
Thus, a simple scoring ratio, i.e., $s_{i,j}=\frac{s_i}{s_i+s_j},$ can result in an invalid strength evaluation.
Therefore, the score of each team is increased by one. 
This modification is known as Colley's method \cite{colley2002colley}, and was originally used to rank college (American) football teams.

This mathematical structure is the well-known {\it logistic regression model}.
It is widely used in areas such as the winning probability assumption of Elo ratings in chess games \cite{EloRating}, 
and the correct answer probability for questions in item response theory \cite{R199109}.

The update method is designed to minimize the sum of the squared error $E^2$ between the result and the prediction , defined by the following equation:
\begin{equation}
\label{eqn:defE2}
E^2=\sum_{(i,j)\in {\mathrm {all~matches}}}(s_{i,j}-p_{i,j})^2.
\end{equation}
It is straightforward to obtain the following update based on the steepest-descent method:
	\begin{equation}
	r_i\leftarrow r_i - \alpha \cdot \frac{\partial E^2}{\partial r_i}, ~~
		r_{homeAdv}\leftarrow r_{homeAdv} - \alpha \cdot \frac{\partial E^2}{\partial r_{homeAdv}},
	\end{equation}
	where $\alpha$ is a constant.

By definition,  the rating is an interval scale.
Therefore, its origin, $r=0$, can be selected arbitrarily, and a constant value can be added to all $r_i$.
For example, 
\begin{equation}
\vec{r}\leftarrow \vec{r}-\left(\max \vec{r}\right)\cdot\vec{1}
\end{equation}
implies that $r=0$ is always the highest rating, whereas $r<0$ is the distance from the top team.

\subsubsection{Conversion of rating on scoring ratio to winning probability}
\label{sec:ratingToWinProb}
The rating $r_i$ in (\ref{eqn:rateDiffToProb}) determines the scoring ratio.
Once we have the scoring ratio $p_{i,j}$ given in (\ref{eqn:rateDiffToProb}), the following independent Bernoulli process is executed $N$ times,  starting from $(s_i, s_j)=(0, 0)$ and with the parameter $0<\beta\leq 1$:
\begin{equation}
\left\{
	\begin{array}{lcl}
	s_i\leftarrow s_i+1 & {\mathrm {with~probability}}&\beta p_{i,j},\\
	s_j\leftarrow s_j+1 & {\mathrm {with~probability}}&\beta\left(1-p_{i,j}\right),\\
	s_i\leftarrow s_i, s_j\leftarrow s_j & {\mathrm {with~probability}}&\left(1-\beta\right).\\
	\end{array}
\right.
\end{equation}
This is a unified (and approximated) model of a scoring process for all ball games, where $s_i$ and $s_j$ model the scores of teams $i$ and $j$, respectively.

{
The parameters $N$ and $\beta$ vary among the sports and between definitions of a unit of play.
For example, in basketball, if a unit of play is defined as 10 [s], we have  $N=40 [{\mathrm{min}}]\times 60 [\mathrm{s/min}]/10 [s]=240$.
In football, if a unit of play is defined as 1 [min], we have $N=90$. 
}
For both sports, $\beta$ is determined to be $\beta=E(s_i+s_j)/N$.

\begin{figure}[h]
\begin{center}
	\includegraphics[width=0.5\columnwidth ,clip]{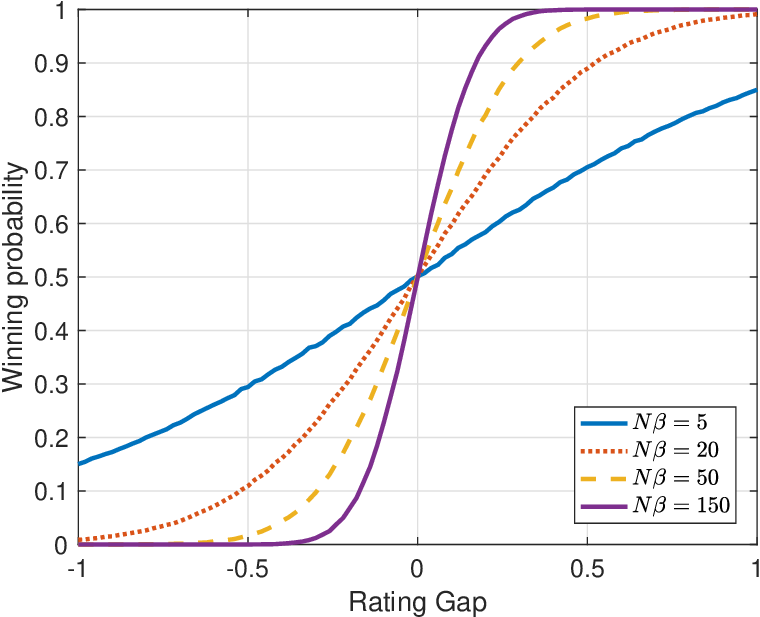}
	\caption{Winning probability with respect to rating gap}
	\label{fig:ratingGapToWinningProbabilitySim}
\end{center}
\end{figure}

{
At the end of the match, $s_i>s_j$ indicates that team $i$ wins against team $j$.
Figure \ref{fig:ratingGapToWinningProbabilitySim} shows the simulated winning probability for different values of $N\beta$ and rating gap ($r_i-r_j$), with $N=240$.
This probability is expressed by as a cumulative distribution function for a normal distribution.
In many applications, it is common to use a logistic regression model rather than a cumulative distribution \cite{10.1504/IJAPR.2013.052339}.
}

{
Based on the previous discussions, we convert the rating on the scoring ratio to that of a winning probability, as follows:}
\begin{equation}
w_{i,j}=1~(i~ \mathrm{wins}),~~ 0.5~(\mathrm{draw}), {\mathrm{ ~~or~~}} 0~ (j~\mathrm{wins}),
\end{equation}
which denotes a  win, draw, or loss, respectively, for team $i$ against team $j$.
Afterward, $D_k^*,$  where $k$ is an index of sports, that satisfies
\begin{equation}
\hat {w}_{i,j}=\frac{1}{1+\exp \left(-D_k\left(r_i+r_{homrAdv}-r_j \right) \right)},
\end{equation}
\begin{equation}
\label{eqn:defDk*}
D_k^*=\arg \min_{D_k} \sum \left(w_{i,j}-\hat{w}_{i,j} \right)^2,
\end{equation}
is obtained. $r_i$ is then converted as follows:
\begin{equation}
\bar{r}_i=D_k^*r_i, ~~ i=1,2,\cdots, N_T, \mathrm{and}~ {homeAdv},
\end{equation}
where $N_T$ denotes the number of teams.
Therefore, $\bar{r}_{homeAdv}$ is a quantitative home advantage estimation that explains the effect on the winning probability.

{
In Equations (\ref{eqn:defE2}) and (\ref{eqn:defDk*}), the sum of squared errors, instead of the cross-entropy, is used as a loss function.
This is because these problems are regression problems, not classification ones.
}

\subsection{Short-term estimation of home advantage}

The proposed method in the previous section was used to estimate the rating of each team and the home advantage in the league for every matchweek using the results of the last five matchweeks, including itself.

By using five matchweeks, we were able to estimate the average of each team's strength and league-wide home advantage over periods ranginf from approximately three weeks to one month.

The calculated home advantage was classified into the following four classes based on  spectator attendance.
\begin{itemize}	
\item Past: The home advantage calculated using the matches from the 2010/11 to  the 2018/19 seasons. This value includes closed matches as punishments, if they exist.
\item Normal: The home advantage calculated using the matches before suspension in the 2019/20 season. These matches all included live spectators. 
Note that closed matches as punishments are also inclused here, if they exist.
\item Mixed:  The home advantage calculated using both the matches that included spectators and those without spectators.
\item Closed: The home advantage calculated using the matches without spectators.
\end{itemize}

\section{Results and discussions}
\label{sec:result}
This chapter describes the analysis results and discussions.

\subsection{Basic statistics}
\begin{figure}[h]
\begin{center}
	\includegraphics[width=0.75\columnwidth]{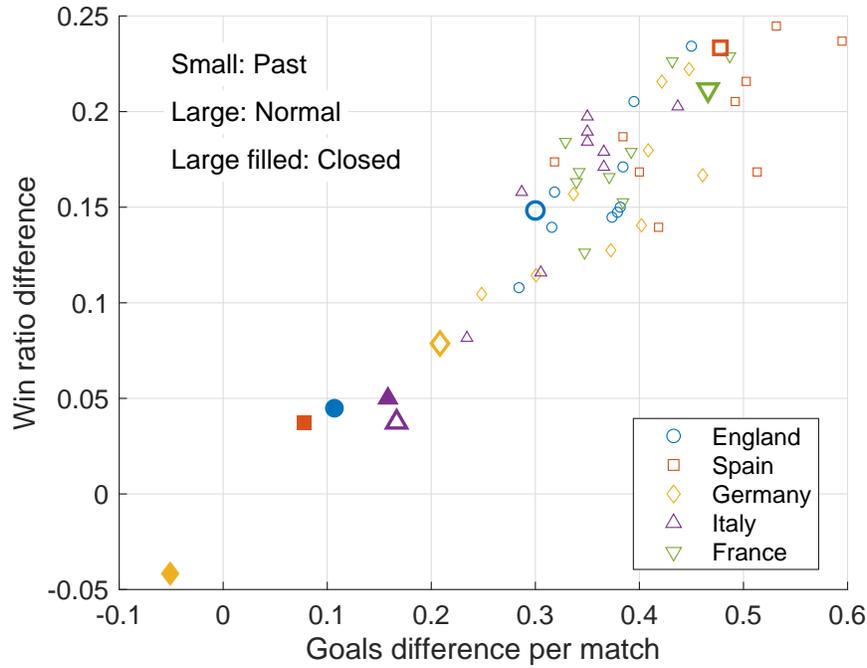}
	\caption{Win ratio difference with respect to goals difference}
	\label{fig:basicStats}
\end{center}
\end{figure}

Figure \ref{fig:basicStats} depicts the basic statistics, e.g., goals difference per match  and win ratio difference, in normal and closed-match periods.

According to these data, home teams goaled approximately from 0.3 to 0.5 more per match than away teams on average under ``past and normal'' situations.
As a result, home teams won approximately 0.15 to 0.25 more against away teams.  
In addition, the home advantage was apparently reduced under closed-match situations. 
In particular, in Bundesliga, the goals difference and win ratio difference were both negative in the closed matches.
In this part of the study, however, possible schedule unbalance for the closed matches was not considered. In other words, it is possible that the most of the home teams were consistently weak (or strong) in the closed matches.

\subsection{Estimation of home advantage}

\begin{figure}[h]
\begin{center}
	\includegraphics[width=0.8\columnwidth]{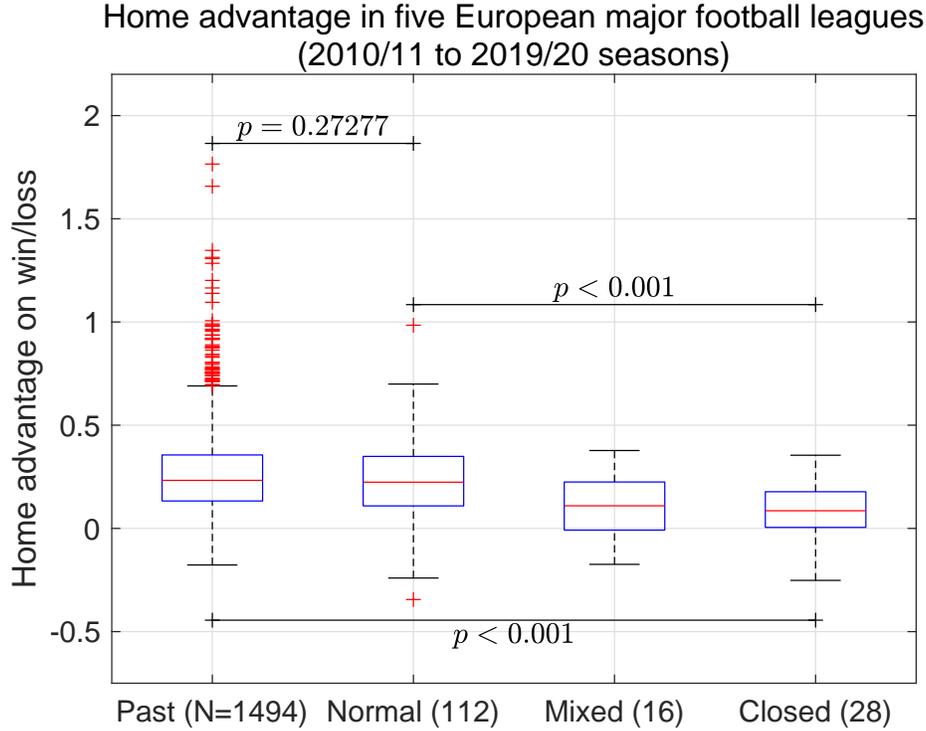}
	\vspace{10pt}
	\caption{Distribution of $\overline{r}_{homeAdv}$ in five major European football leagues}
	\label{fig:homeAdvIn5Majors}
\end{center}
\end{figure}

Figure \ref{fig:homeAdvIn5Majors} shows the results of the estimation of home advantage $\bar{r}_{homeAdv}$ for five leagues.
The medians were all positive for every four classes. 
This result indicates that the home advantage remained even for matches that were closed and without spectators.
The medians in the past and normal periods appeared similar.
On the other hand, the median in the closed period was smaller than those of the past and  normal  periods.

We tested a null hypothesis that the home advantage $\bar{r}_{homeAdv}$ from two different categories were samples from continuous distributions with equal medians.
Wilcoxon's rank sum test \cite{ranksumMATLAB,gibbons2010nonparametric} was used as a test method because any assumption on the shape of the distribution of $\bar{r}_{homeAdv}$ could be posed.
The $p$-values between classes are depicted in Figure  \ref{fig:homeAdvIn5Majors}, whereas
the test results are summarized in Table \ref{tab:hypoTest}.

\begin{table}[htbp]
\begin{center}
\caption{Test results: Overall}
\label{tab:hypoTest}
\begin{tabular}[t]{llcccrr}\hline 
Sample X & Sample Y & $N_X$ & $N_Y$ & $p$-value & $z$-value & ranksum \\ \hline
Past & Normal & 1494 & 112 & $2.72\times 10^{-1}$ & 1.097 & $1.207\times 10^{6}$ \\
Past & Mixed & 1494 & 16 & $3.27\times 10^{-3}$ & 2.941 & $1.134\times 10^{6}$ \\
Past & Closed & 1494 & 28 & $5.44\times 10^{-7}$ & 5.010 & $1.149\times 10^{6}$ \\
Normal & Mixed & 112 & 16 & $2.72\times 10^{-2}$ & 2.208 & 7531 \\
Normal & Closed & 112 & 28 & $1.98\times 10^{-4}$ & 3.722 & 8611 \\
Mixed & Closed & 16 & 28 & $5.34\times 10^{-1}$ & 0.622 & 386 \\ \hline
\end{tabular}
\end{center}
\end{table}

From these test results, the following can be concluded.
\begin{itemize}
\item There was no significant difference in home advantage between the past's and 2019/20 season's normal matches ($p>0.1$).
\item There was significant difference in home advantage between the normal and closed matches $(p<10^{-3})$.
The significantnce of the difference was even more obvious between the past and closed matches ($p<10^{-6}$).
The median of the home advantage in the closed matches was clearly smaller.
\end{itemize}

Therefore, this study provides strong quantitative evidence of the impact of the crowd effect on home advantage among the top leagues in Europe.
It should be noted, however, that all of the closed matches in this season have been with  several months' suspension, and conducted in the summer, when matches are normally not  played.
Therefore, the possible effect of these factors  on the home advantage should not be neglected.

\subsubsection{Detailed analysis for each league}
\begin{figure}[h]
\begin{center}
	\includegraphics[width=0.75\columnwidth]{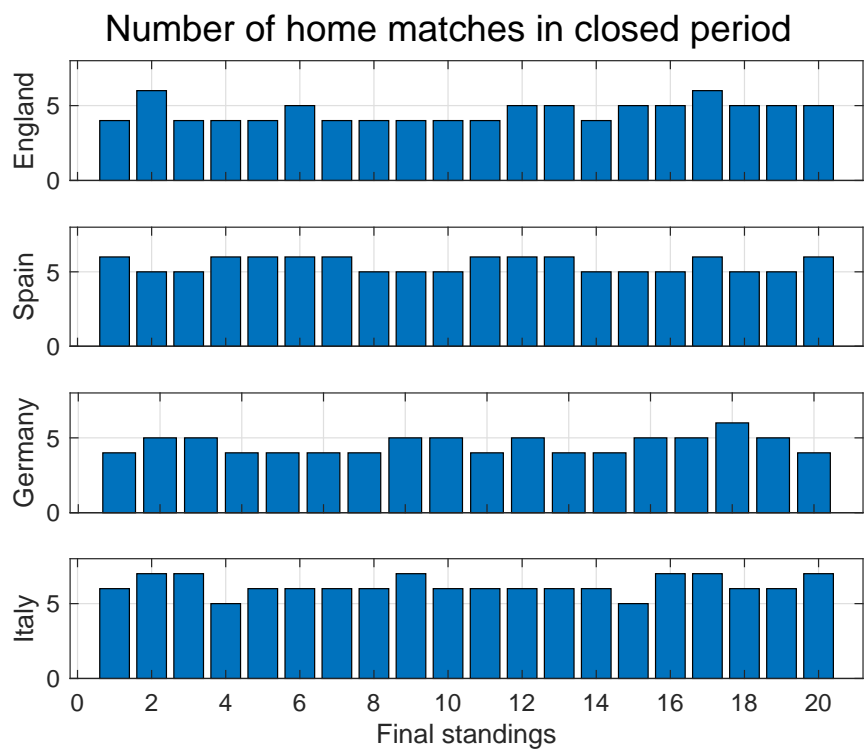}
	\caption{Number of home matches in closed-match period}
	\label{fig:numOfHomeMatchesInClosedPeriod}
\end{center}
\end{figure}

\begin{figure}[h]
\begin{center}
	\includegraphics[width=0.75\columnwidth]{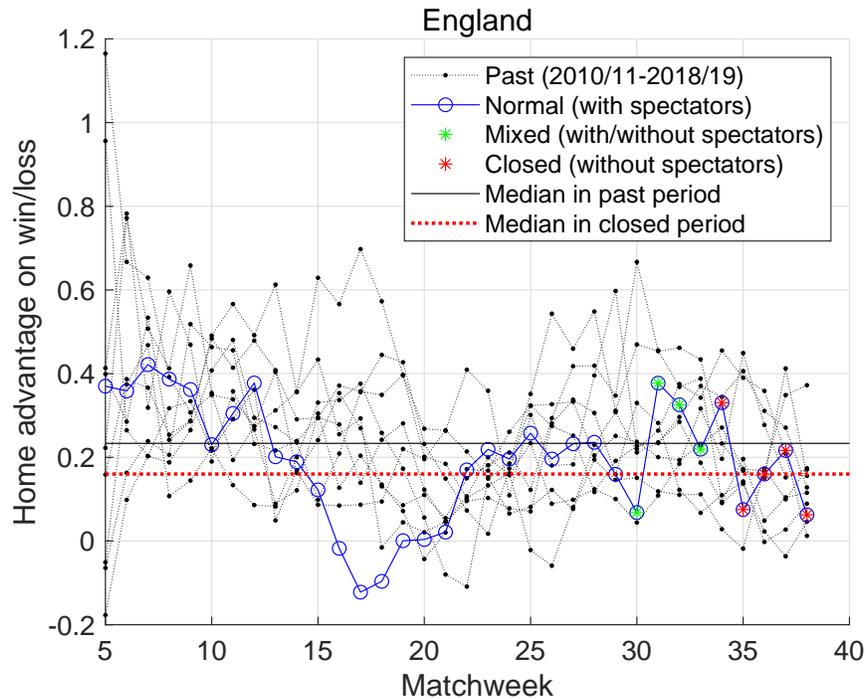}
	\caption{History of $\bar{r}_{homeAdv}$: England}
	\label{fig:homeAdvInEngland}
\end{center}
\end{figure}

\begin{figure}[h]
\begin{center}
	\includegraphics[width=0.75\columnwidth]{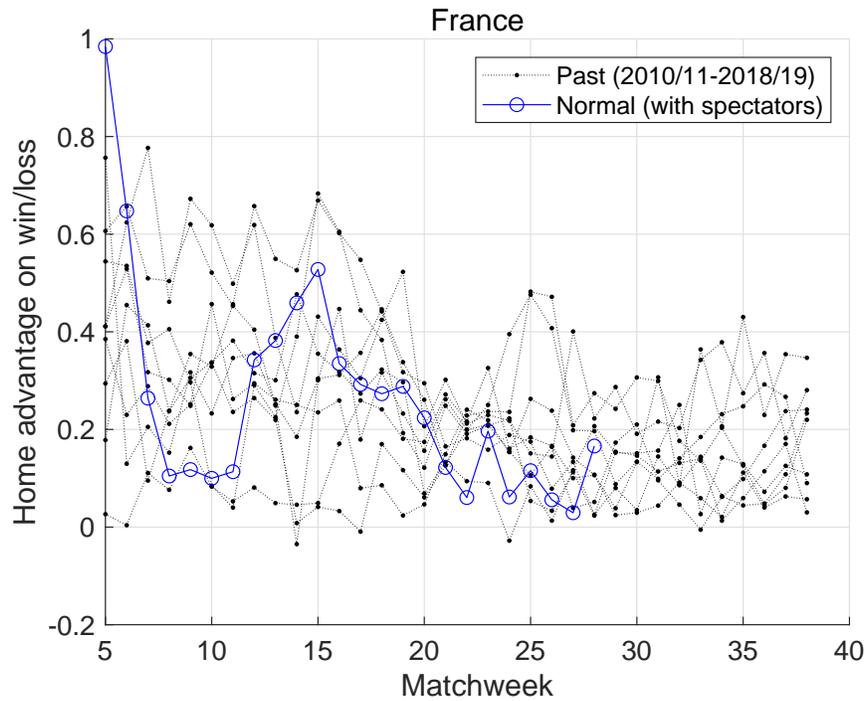}
	\caption{History of $\bar{r}_{homeAdv}$: France}
	\label{fig:homeAdvInFrance}
\end{center}
\end{figure}

\begin{figure}[h]
\begin{center}
	\includegraphics[width=0.75\columnwidth]{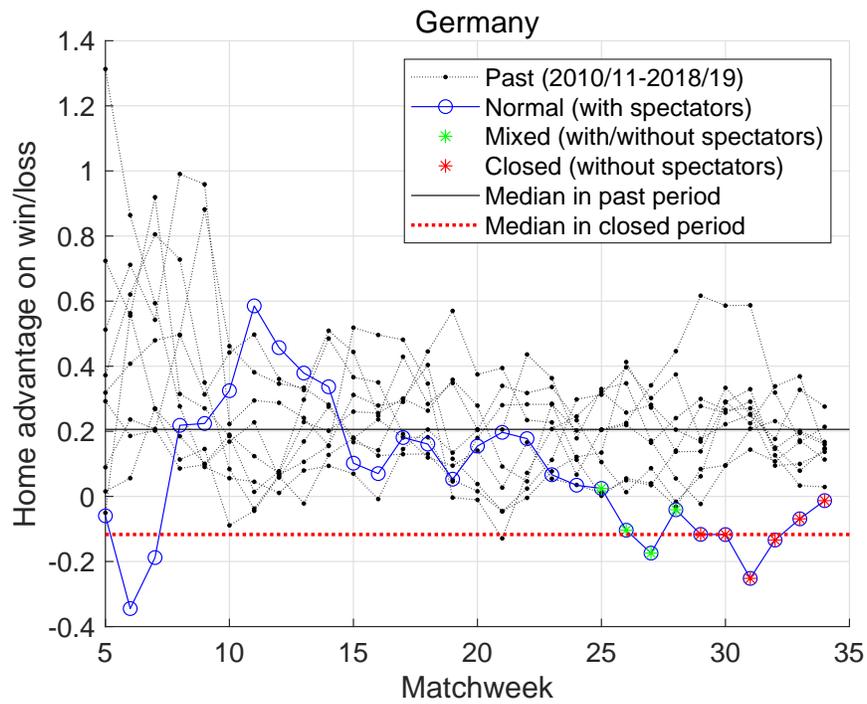}
	\caption{History of $\bar{r}_{homeAdv}$: Germany}
	\label{fig:homeAdvInGermany}
\end{center}
\end{figure}

\begin{figure}[h]
\begin{center}
	\includegraphics[width=0.75\columnwidth]{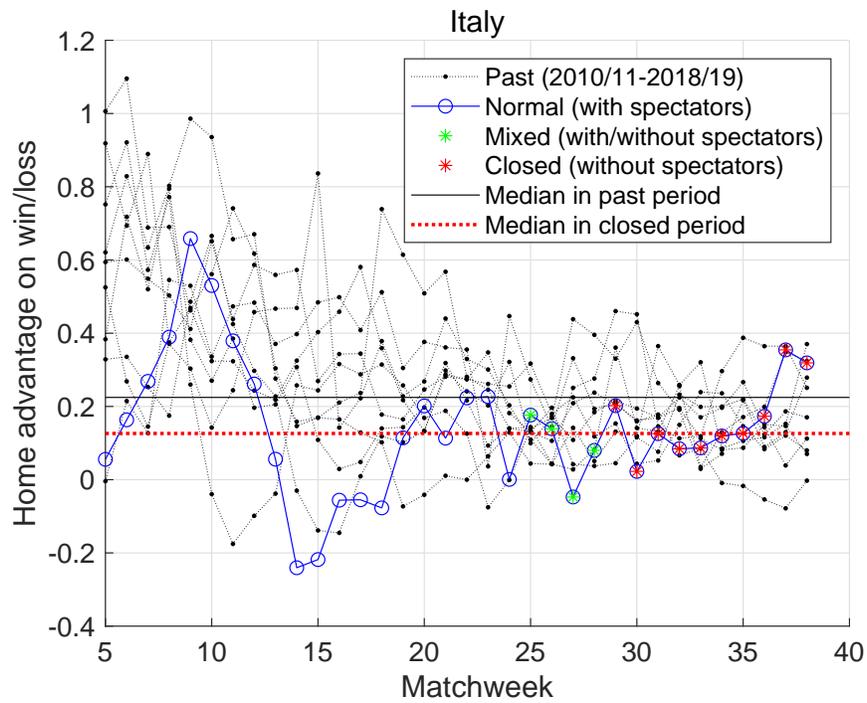}
	\caption{History of $\bar{r}_{homeAdv}$: Italy}
	\label{fig:homeAdvInItaly}
\end{center}
\end{figure}

\begin{figure}[h]
\begin{center}
	\includegraphics[width=0.75\columnwidth]{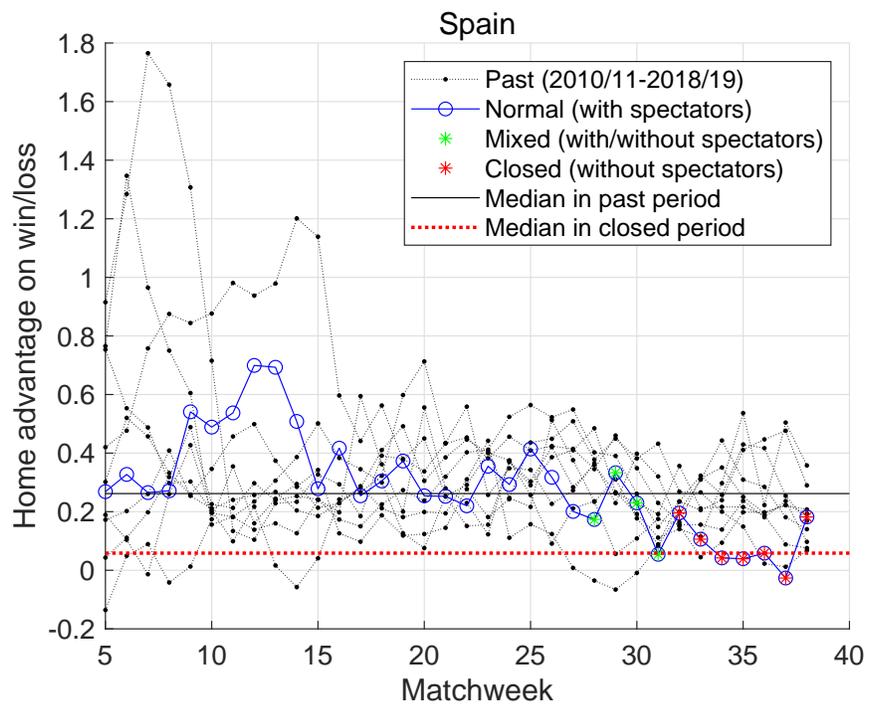}
	\caption{History of $\bar{r}_{homeAdv}$: Spain}
	\label{fig:homeAdvInSpain}
\end{center}
\end{figure}

The previous section described the results for all five leagues.
In this section, the estimation results for each league will be explained.
Table \ref{tab:homeAdvInEachLeague} lists the results of Wilcoxon's rank sum test for each league in detail. 
Figure \ref{fig:numOfHomeMatchesInClosedPeriod} visualizes the number of home matches in the closed-match period. The teams are ordered based on their the final standings.
Figures \ref{fig:homeAdvInEngland} to \ref{fig:homeAdvInSpain} show the estimated value of $\overline{r}_{homeAdv}$ for each league from the 2010/11 season.

\begin{table}[h]
\begin{center}
\caption{Test results: League breakdown}
\label{tab:homeAdvInEachLeague}
\begin{tabular}[t]{llccccrr}\hline 
League & Sample X & Sample Y & $N_X$ & $N_Y$ & $p$-value & $z$-
value & ranksum \\ \hline
England & Past & Normal & 306 & 25 & $2.23\times 10^{-1}$ & $-1.218$  & 3589 \\
England & Past & Closed & 306 & 5 & $2.13\times 10^{-1}$ & 1.246  & 47985 \\
England & Normal & Closed & 25 & 5 & $5.78\times 10^{-1}$ & $0.556$  & 398 \\

France & Past & Normal & 306 & 24 & $9.35\times 10^{-1}$ & $0.081$  & 4009
 \\
 Germany & Past & Normal & 270 & 20 & $1.63\times 10^{-1}$ & $-1.394$  & 2405 \\
Germany & Past & Closed & 270 & 6 & $3.98\times 10^{-5}$ & 4.109  & 38190 \\
Germany & Normal & Closed & 20 & 6 & $3.84\times 10^{-3}$ & 2.891  & 318 \\

Italy & Past & Normal & 306 & 20 & $2.85\times 10^{-2}$ & $-2.190$  & 2375 \\
Italy & Past & Closed & 306 & 10 & $6.46\times 10^{-2}$ & 1.848  & 49027 \\
Italy & Normal & Closed & 20 & 10 & $7.75\times 10^{-1}$ & $-0.286$  & 303 \\

Spain & Past & Normal & 306 & 23 & $2.06\times 10^{-2}$ & 2.315  & 4814 \\
Spain & Past & Closed & 306 & 7 & $4.10\times 10^{-4}$ & 3.533  & 48879 \\
Spain & Normal & Closed & 23 & 7 & $8.76\times 10^{-5} $& 3.923  & 437 \\
 \hline
\end{tabular}
\end{center}
\end{table}

\flushleft{\bf England}
In England, the home advantage value obtained via the proposed method  for the closed-match period had no significant difference to those of  the past and normal periods.
This result demonstrates that the basic statistics, such as the goals difference and win ratio difference shown in Figure \ref{fig:basicStats}, were smaller in the closed-match period because of the unbalanced schedule. 
In other words, most of the home teams were consistently weak (or strong) in the closed matches.
Figure \ref{fig:numOfHomeMatchesInClosedPeriod} supports this assertion.
In England, there were weak correlation between the number of home matches in the closed-match period and the final standings. The correlation value was $0.4052$.
By constrast, for the other leagues, i.e., Germany, Italy, and Spain, the correlation values were much smaller ($0.2863$, $0.0867$, and $-0.1561$).
 
\flushleft{\bf France}
In France, there was not a significant difference in $\overline{r}_{homeAdv}$ between the past and normal periods.

\flushleft{\bf Germany}
In Germany, the home advantage value in the closed-match period had significant difference to those  in the past and normal periods. Both $p$-values were less than $1.00\times10^{-2}$.
In the closed-match period, Bundesliga demonstrated a home ``dis''advantage, i.e., $\overline{r}_{homeAdv}<0$.

\flushleft{\bf Italy}
In Italy, the home advantage value obtained via the proposed method  for the closed-match period had no significant difference to those of  the past period ($p=6.46\times 10^{-2}$).
In this season, the home advantage value was smaller even in the normal period.

\flushleft{\bf Spain}
In Spain, the home advantage value in the closed-match period had significant difference to those in the past and normal periods. Both $p$-values were less than $1.00\times10^{-2}$.
In the closed-match period, similar to Bundesliga, LaLiga demonstrated home disadvantage.

\section{Conclusion}
\label{sec:conclusion}
In this study, the results of matches conducted behind closed doors during the COVID-19 pandemic were used to determine the relationship between the presence of spectators and home advantage.
To reduce the effect of schedule unbalance in the closed-match period, this paper proposed a short-term (e.g., five matchweeks) rating method that considers the home advantage.
The proposed method was applied to the match results in five major European football leagues (England, France, Germany, Italy, and Spain) from the 2010/11 to  the 2019/20 seasons.

The distributions of home advantage for both the past normal and closed-match periods were calculated.
Their median values were compared using statistical hypothesis tests.
A null hypothesis, ``the home advantage $\bar{r}_{homeAdv}$ from two different periods are samples from continuous distributions with equal medians,'' were rejected because of  sufficiently small $p$-value ($p<10^{-3}$). 
More simply, the home advantage became smaller when the games were conducted behind closed doors

Our future work is to extend the proposed method to match results from all over the world.
This future study could then clarify the crowd effect on home advantage.


\bibliographystyle{unsrt}


\end{document}